# WiFi-based Crowd Monitoring and Workspace Planning for COVID-19 Recovery

Mu Mu, University of Northampton, Northampton, NN1 5PH, United Kingdom

## Abstract

The recovery phase of the COVID-19 pandemic requires careful planning and monitoring while people gradually return to work. Internet-of-Things (IoT) is widely regarded as a crucial tool to help combating COVID-19 pandemic in many areas and societies. In particular, the heterogeneous data captured by IoT solutions can inform policy making and quick responses to community events. This article introduces a novel IoT crowd monitoring solution which uses software defined networks (SDN) assisted WiFi access points as 24/7 sensors to monitor and analyze the use of physical space. Prototypes and crowd behavior models are developed using over 500 million records captured on a university campus. Besides supporting informed decision at institution level, the results can be used by individual visitors to plan or schedule their access to facilities.

## Introduction

Many countries have past the peak of the coronavirus outbreak and now set to restart their economy. One of the main objectives at the recovery stage is to help workers, businesses, and organisations establishing safe social interactions and work practices. Many business organisations are looking for a local crowd monitoring solution to evaluate how different parts of their premises are used by employees or customers so that risk assessment and informed planning can be made to assist social distancing. Meanwhile, nation-wide or global ICT solutions such as contact-tracing mobile apps are being developed and tested to play a central role in managing the rate of COVID-19 transmission. Most contact-tracing solutions use Bluetooth-based proximity events to estimate infection risk and trigger notifications [1]. Such solutions require app installation and there is no clear pathway for the apps to directly benefit the operations at individual organisations due to the lack of access to tracing data. Some other crowd monitoring solutions use audio-visual object detection or LiDAR to monitor social distancing and crowd [2] [3]. They are highly dependent on specialised equipment like surveillance cameras and high-resolution sensors which can be costly and intrusive to everyday life.

This article introduces a WiFi-based solution that exploits data readily available in an existing WiFi infrastructure. It is not dependent on a large capital investment or installation on user devices. Our design uses WiFi access points (APs) as a network of always-on sensors. It combines the analysis of WiFi roaming patterns and geospatial information of APs to enable real-time monitoring and predictive analysis of crowd movements in physical space. This data-driven approach will allow organisations to reduce unnecessary human-to-human contact through the improved workforce scheduling and space allocation. Individual visitors can use a user facing application to check crowd density and plan their trip to work or a business site. The solution is built upon a smart campus project which has already established a live dataset of over 500 million WiFi records to model crowd behaviours. The remainder of the paper introduces the supporting systems and a range of applications developed to help combating COVID-19.

## System overview

Figure 1 illustrates the WiFi-based crowd monitoring system deployed at a University campus. The campus has a state-of-the-art software-defined network (SDN) infrastructure to provide internet connectivity to its residence and visitors. SDN champions the new network paradigm that separates network control from the data forwarding functions (hardware equipment). This allows network equipment from different vendors to be managed by a single network controller as long as they conform to the same SDN specifications. 1,500 indoor and outdoor APs provide wireless internet access to mobile user devices across the campus. Each AP connects to a network switch via a wired connection and projects Wi-Fi signal to cover a designated area. Extended WiFi coverage is achieved by strategically placing multiple APs. As user devices make location changes, they automatically detect the APs available in the area and switch between APs seamlessly based on signal strength to each AP.

On the campus where our designs and testing were made, most buildings on site have between 20 to 55 APs installed on each floor depending on the functions and the internal structures of the building. Most rooms have at least one dedicated AP while some large rooms, foyers and corridors have several APs to warrant an excellent WiFi coverage. There are over 30 outdoor APs for people who use outdoor workspace and for the continuous internet connectivity while visitor travel between buildings. The physical location and installation mode (e.g., ceiling, wall, post) of each AP was recorded for maintenance purposes. For instance, one AP is marked on the floor map with the descriptive text "outside room 210" and its installation type is "ceiling". With SDN, a single network controller manages all network switches and wireless APs. This enables a live and unified view of all devices connected to each AP. Overall, the networking designs adopted the same design principles adopted by other organizations. Our solution uses exiting information readily available at network controller(s) and does not require changes in the network.

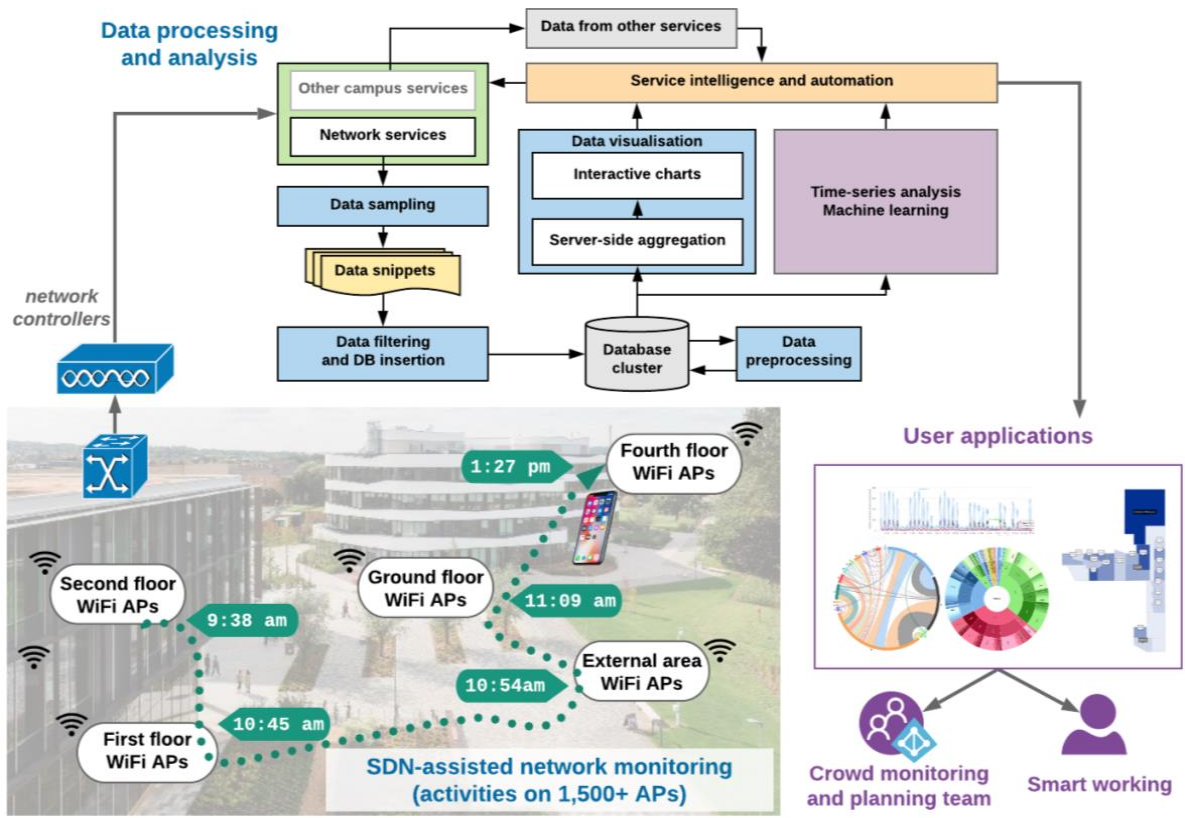

*Figure 1 smart campus system*

In order to study the crowd dynamics over time and to support predictive analysis, live device information at each AP are sampled and stored. Each sample is a snapshot of all wirelessly connected devices and their associated APs. The number of devices is usually in the range of 1,000 to 6,000 depending on the time of the day and the day of the week. A dedicated server takes one sample every 60 seconds. A data processing function goes through all unprocessed samples and populates a database asynchronously with filtered and formatted data. The sampling rate was chosen based on four factors: how frequent do devices switch between APs, the balance between noise and fidelity, the impact to the network controller APIs, and the storage capacity. Overall the data collection process gathers over one million records per day. The raw data include over 20 data fields such as *datetime*, *MAC address*, *IP address*, *WiFi SSID*, *user ID*, *device type*, and *WiFi protocol*. Visitors may have more than one device connected with the same user ID. Associating time-coded records of connected devices and the geo-spatial information of APs, the system can discover how connected devices are distributed on site and how their locations change over time. Since July 2019, over 500 million records of data have been collected. Table 1 lists modified example records based on the raw data captured from network controllers.

*Table 1 Illustration of data captured from network controllers*

| date time | MAC address | IP address | AP name | network name | user ID |
|---|---|---|---|---|---|
| 02/07/2020 10:00:01 | b6773f96309c5aa22a81ee034ff7f559 | a77803ead788503528d8c3b7f4650798 | AP-MB-073aca | UoN_Guest | b475a86b6789657cb91aec5d35199e1f |
| 02/07/2020 10:00:01 | c6a54e36752ad3fb8bdd8789fc48f78a | eddc6c827a4efd013f3eb2d8c83da70f | AP-PA-ce6243 | UoN_Student | b07d4a81b42a55843ad90372891afdd4 |
| 02/07/2020 10:00:01 | a3bd6f7c930955aa8041006a87822c6c | 2e5df82147720bd9a6c06341085e5163 | AP-PA-a7a404 | UoN_Student | b766002b992e6b9126a860c121322764 |
| 02/07/2020 10:00:01 | a1f4da2723d3abeeb4ffdea5725fe9fe | 9dc8161de5d5db9f2128d44cf7a1ca09 | AP-FC-be6d34 | eduroam | 9fd88b12be4e29077c812f67f1298563 |
| 02/07/2020 10:00:01 | f3952fc8b937e46ccb2c546809000c7a | 3e5e5cf2a7d91a60c59a52694cdbafaf | AP-SP-e3d04e | SSN | 0ca55239ec92e00a8f241827aab87197 |
| 02/07/2020 10:00:01 | a24a7eb47cb3fc0f2cf1894b471c61ad | ba1547ba8a3d2c137f53a3437bd97d00 | AP-SP-e3d04e | SSN | 0ca55239ec92e00a8f241827aab87197 |
| 02/07/2020 10:00:01 | e6b3f04847a3d6ec6ca0838efb17344c | e3fe301f433c36584f4144ee71619e16 | AP-LH-b71103 | UoN_Guest | 1bab53c8e7701ae19cdf9c0600b958b0 |
| 02/07/2020 10:00:01 | 553fe2b11b0276ff3e56606775bd17d5 | 5a72482963b47052cf199385e92b74f7 | AP-EX-4c6223 | UoN_Staff | 7a2db8b5da9df1af718dac328aa82ab2 |
| 02/07/2020 10:00:01 | bf340ebab92ffd058fbca491fe8a810e | 1498188b8e522b3f1b2f3aced830b29c | AP-LH-b71103 | UoN_Staff | afd57ae92e3bb268a44bcba932f501d0 |
| 02/07/2020 10:00:01 | 92d281f900caea374c36840222d6c403 | 43cb1fb6336fa16bf1f670db4f6a2610 | AP-PA-ca35c8 | UoN_Student | 3eb93fd0c0fc0e3af148cec22b64de0c |
| 02/07/2020 10:00:01 | 3438b08e88caa00e74ff0b2cd49573af | 039b663892529e2e9f63866df08d24e3 | AP-FC-c55812 | UoN_Gaming | 3f6e9a7cb490f8451ed4e25644d6edde |
| 02/07/2020 10:00:01 | 2d58639d0ebf3c6b2f65ac60625bf1af | eea5d4f7d98fa3444323fc8942a5ef53 | AP-EX-e74152 | UoN_Guest | c1221a0cc64bb3cc3afc83563ff40cc4 |
| 02/07/2020 10:00:01 | 8f2629130314115ab3e3f19fc232fd07 | ca0a2f88b18695fe964b10f8da7f47bb | AP-MB-2f9022 | UoN_Student | 2fc002cecbd93ed4a25bbe2cf9a34887 |
| 02/07/2020 10:00:01 | c867421c3c34d10214f68e70cf4eb37c | a2b7c90f6021c5cdc6458853ecee7f7d | AP-LG-e381d8 | eduroam | e15828712cecc38830ee99c6a35cc1be |
| 02/07/2020 10:00:01 | b9f66a7c0afb2e365d371e0f0c02045a | 3556a889d95f627bda5bf529fb0d87b6 | AP-PA-227f03 | UoN_Staff | e917a1d584bc134b96923bb59cdaa167 |
| 02/07/2020 10:00:01 | 16bbe07b42320402b688545c248c7f29 | e06227e4476ea0b2ff1203c0102d0621 | AP-PA-fed0cd | UoN_Guest | 0ca55239ec92e00a8f241827aab87197 |
| 02/07/2020 10:00:01 | c1339963cc359bda370cbc09adbec732 | bbda7c686b458b756ff8d268ff39129b | AP-PA-5ed3b6 | UoN_Student | f7939a5171db387f7c525ef7944bb4f2 |
| 02/07/2020 10:00:01 | e0dfc56530be09297b24a5cb4d6ddf11 | 8d02661662fcd564105a2189b78a9711 | AP-CB-bb81e8 | UoN_Student | eb9b08c7d9e9c3fd3ab59204e75575a8 |
| 02/07/2020 10:00:01 | 75e15c13d32bac8493184ee25e597dd1 | 1f1635aecbb75b6186365b9e35072e25 | AP-CB-bb81e8 | UoN_Student | 1dc108de42b6f0c11a4853db319e8c59 |

Personal and device information are anonymized via a custom hashing function as part of the sampling process. Due to the low probability of hash collision, the hashed data are considered unique and distinguishable. Hence, we can associate anonymised device IDs and user IDs within each sample and across multiple samples to estimate the number of visitors (unique user IDs) in an area (defined by the coverage of one or multiple APs) and how each person moves between areas.

A range of data analysis and visualization applications have been developed to extract high-level information from the large volume of data collected. The aim of the applications is to inform decision making at both organisational and individual visitor level. The original use cases of the project were promoting social interactions and closeness, energy efficiency, safety, and timetabling. To support COVID-19 recovery, many features have been redesigned to support safeguarding and social distancing. Examples of the COVID-19 use cases are crowd density risk assessment, pro-active workspace planning and scheduling, event simulation, and lone worker support. The following sections discuss the modelling over behaviour data and developments made to support the aforementioned use cases.

## Crowd dynamics

Figure 2 depicts the number of campus visitors (both indoor and outdoor) between September 2019 and May 2020 as well as the detailed floor-level data for one particular building during the three-week period leading to the COVID-19 lockdown in the UK. The figure shows two distinctive clusters of data corresponding to the two university terms separated by the end of the year holiday period. Furthermore, there are distinctive weekly and daily seasonal patterns which reflect how students and staff access campus facilities. Most occupants during the weekends were residences of student accommodations on campus. On the weekdays, the number of visitors steadily increased from early morning and peaked at mid-day. Sports activities and events are often scheduled on Wednesday afternoons, which led to less visitors and user devices on Wednesdays.

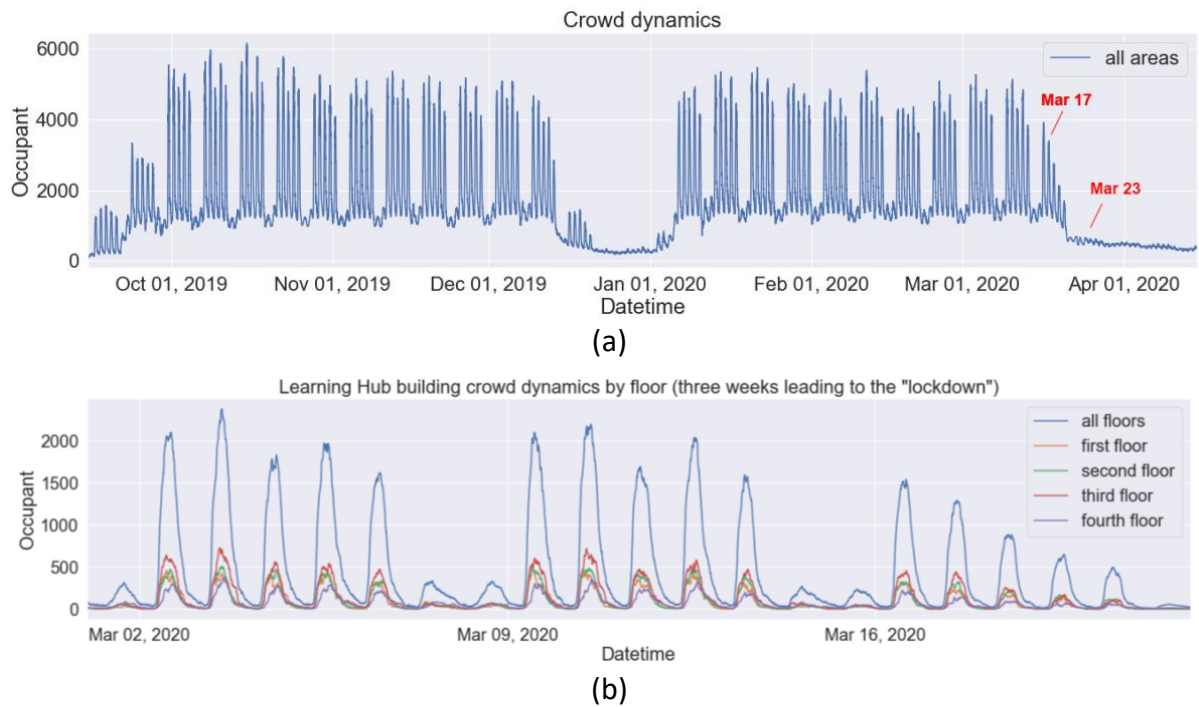

*Figure 2 Crowd dynamics*

The COVID-19 lockdown in the UK has a clear impact to the higher education as it shows in Figure 2. Up until the middle of March our data show regular patterns of access to the campus in spite of the substantial media coverage of COVID-19. Staff came to work as usual and students attended their normal time-tabled lectures on campus. On Tuesday March 17 the university announced plans to close campus on Friday March 20. Although the campus was still fully open during that week when the announcement was made, we observe a clear downward trend of physical access to campus. Students were still actively engaging with learning, but many chose to participate classes online and study at home rather than working in groups using workspace on campus. The UK government imposed the lockdown on the evening of March 23, 2020 [4]. On campus, access to most buildings was prohibited so the substantial increase and decrease of visitor numbers normally observed in the mornings and evenings disappeared. While there were still daily and weekly data patterns contributed by the residence remained on campus, the number continued to diminish while many student residences started to leave for home.

## Crowd modelling and prediction

The capabilities to predict the crowd density at a particular point in time based on historical observations are essential for proactive planning and anomaly detection. Our crowd data are in the form of time series as they are sequences of numerical data points in successive order observed at regular intervals. Time-series analysis enables us to model trends and patterns which can support prediction of the value at a future point. ARIMA (Auto Regressive Integrated Moving Average) [4] is a common time-series analysis method characterised by 3 terms: *p*, *d*, *q* that capture the pattern of changes in the data ("auto-regressive"), the rate of changes in the data ("integrated") and the noise between consecutive time points ("moving average"). The seasonal variation of ARIMA (SARIMA) introduces additional terms to capture seasonal differences for non-stationary data [6]. Our data are non-stationary with multiple levels of seasonality embedded. Taking a top down view: the university organizes most activities in terms, each term breaks into several weeks with special events at the beginning and the end, each week has weekdays and weekends with both lecturing hours and non-lecturing hours on weekdays. Using different season configuration, the SARIMA can capture such seasonality at different levels.

We first used the SARIMA method to predict crowd level on a single day. Hence, the seasonal term was configured to cover observations from a day. Two types of predictions were tested: intra-week prediction and inter-week prediction. The intra-week prediction uses data observed on previous days of a week to conduct estimations for future days in that week, e.g., using data on Monday-Thursday to estimate the data on Friday. The inter-week prediction is based on the data associated with the same weekday but from previous weeks, e.g., using Fridays from previous weeks to predict data on future Fridays.

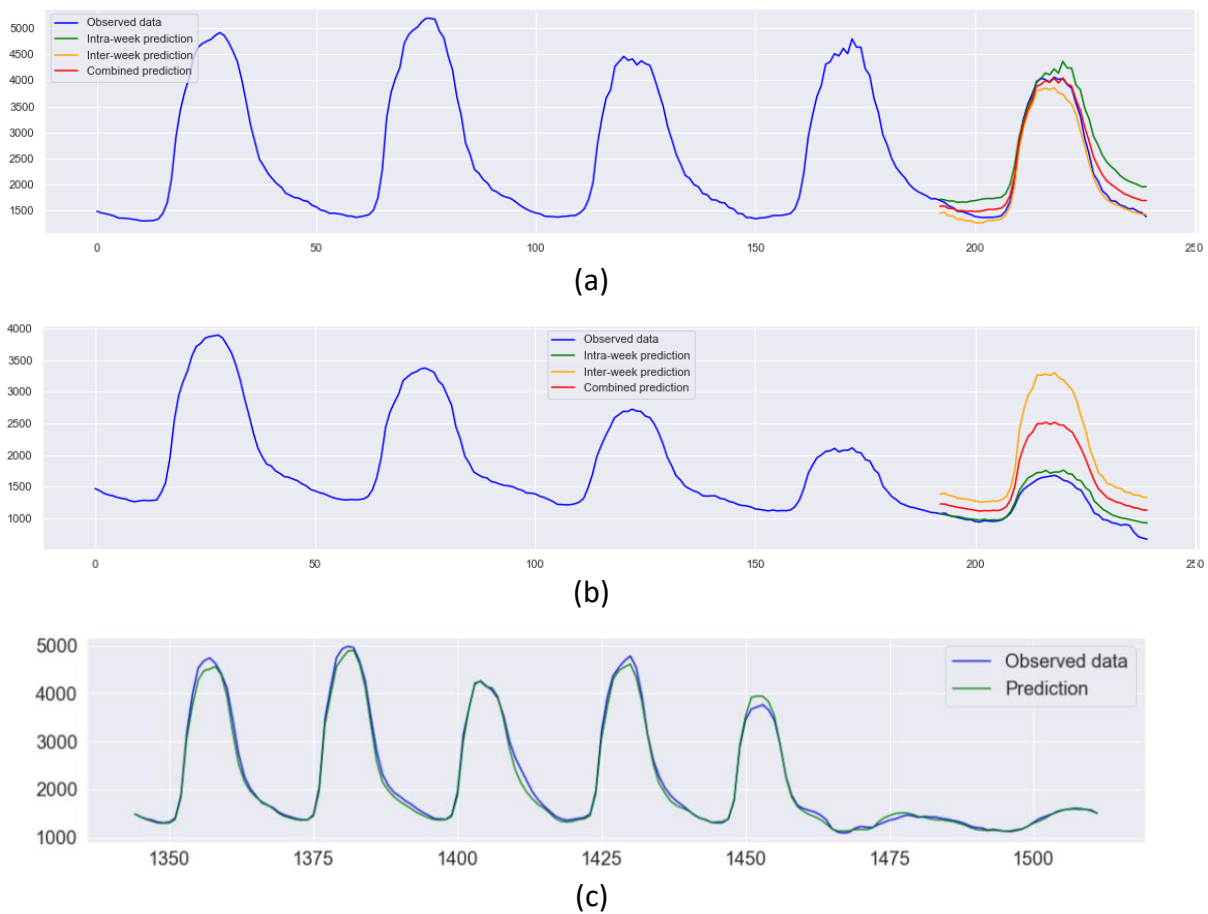

*Figure 3 SARIMA predictions*

Figure 3(a) shows the predictions of the crowd level on Friday March 6 of a normal term week. The blue curve plots the observed data on that week (Monday to Friday). The green curve shows the results of an intra-week prediction where data from four days of the week was modelled. The intra-week prediction slightly overestimates the crowd level for Friday afternoon. This is due to the fact that people normally leave early on Friday afternoon compared with other weekdays. The intra-week prediction cannot capture that extraordinary weekly pattern. The orange curve gives the inter-week prediction results based on previous four Fridays. This method captures normal activities on Fridays but is agnostic to week-specific changes (e.g., the week prior to exams). On this particular Friday, staff and student ambassadors were doing additional work preparing for a "Discovery Day" for prospective students. This explains the slight under-estimation made by the inter-week prediction. Balancing intra- and inter-week predictions using a simple element-wise averaging function, the red curve shows the results when the outcomes from the two predictions are combined with equal weights.

To investigate how the SARIMA-based models respond to COVID-19 induced impact on crowd density, the second test chose the week prior to the COVID-19 lockdown (Figure 3(b)). This week is considered an "abnormal" week as crowd behaviors deviate from normal work patterns whilst people spent more time study or work from home. In this case, the intra-week model successfully captures the gradual day-by-day decrease of visitor numbers in that week, when other methods generally over-estimate the crowd based on previous weeks.

By adjusting the seasonal configuration, the SARIMA method can model and predict data in unit of a week. Using data from 7 consecutive term-time weeks, a model is constructed to predict crowd level of the following week. Training data were down-sampled as 30-minute intervals to speed up the modelling process. Figure 3(c) depicts the results of the week-level prediction. The model is able to capture the unique characteristics of the different days of the week.

## Cross-area movements

Crowd movements indicate how visitors travel between different areas to access services. Such movements are often determined by how physical facilities were designed and how events such as meetings and classes are planned. In some cases, the cross-area movements are encouraged for the benefit of physical and mental health or to facilitate social interactions. There are also scenarios where extensive traveling between areas are considered as a sign of poor planning and scheduling. For instance, simulations of crowd movements are often carried out to evaluate whether a public venue has a sufficient number of access routes to allow a large number of visitors (flash crowd) safely arriving or leaving at the same time. While combating COVID-19, unnecessary movements between floors or buildings may be discouraged as there are often "bottlenecks" such as stairs, elevators, and building entrances when people move between areas. Social distancing is less likely to be kept at the "bottlenecks". Therefore, monitoring and understanding cross-area movements can assist risk assessment at high risk locations.

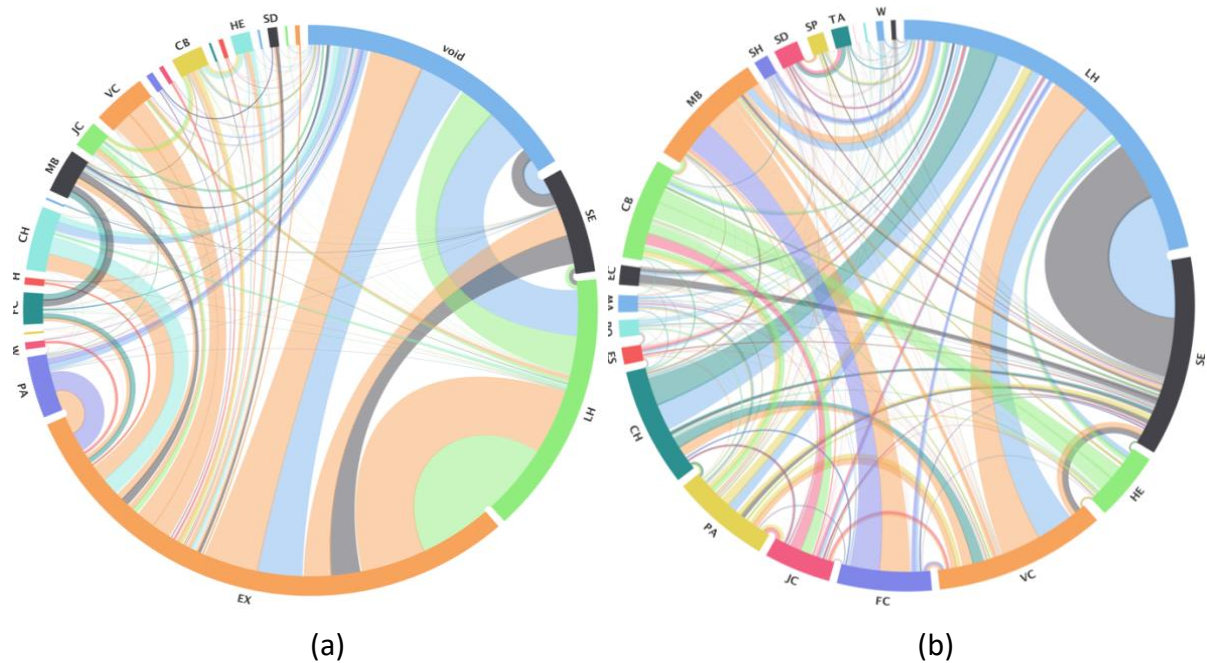

*Figure 4 A dependency graph showing crowd movements*

We developed data analysis functions to examine cross-area location change taken by visitors. Figure 4 (a) shows a dependency graph of how connected devices move between areas and buildings during visiting sessions. A session begins when a device is first seen after at least 2 hours of absence and ends after the device has not been connected to any AP for 2 hours. The short codes at the edge of the graph represent different areas. For instance, "LH" stands for Learning Hub, a building whose data were used in Figure 2. "EX" is the code for all outdoor areas covered by ourdoor WiFi APs. "void" is a pseudo area code used to represent any out-of-campus areas. It helps visualizing where devices were first observed. Normally, a mobile device with WiFi turned on would first connect to outdoor APs ("EX") as visitors approach the campus from car parks or foot bridges. This is illustrated by the light blue outward link from "void" to "EX". The graph also shows many devices first discovered in a building without pass through outdoor APs such as the light blue outward link from "void" to "LH". This is believed to be from laptop PCs which were not switched on until they reached a building, or mobile devices that were not connected to WiFi until they were manually connected inside a building. "EX" bridged most cross-building movements as visitors travelled between different parts of the campus. In order to illustrate how buildings are associated as part of the visitor's journeys, Figure 4 (b) uses direct links between buildings and bypasses the "EX" and the "void". Movements between buildings were part of the campus design to improve the usage of facilities and to facilitate social interactions between people with different area expertise. To support COVID-19 social distance and safe return to work, the interactions between crowd will be managed differently. The dependency graphs shown in Figure 4 will support the evaluation of any temporary arrangement made due to COVID-19.

The 6 million cross-building movement records that underpinned Figure 4 can be modelled using a Markov model based on the assumption that the next area to be visited is determined by where a visitor is currently and not where this visitor was before. It will also be possible to model the movements using artificial neural network models such as recurrent neural network (RNN), which may factor in multiple previous areas for prediction. A potential use case of such modelling is crowd simulation for planning and risk assessment.

## Crowd density heatmaps

For work planning and scheduling, it is essential to monitor and evaluate how facilities such as meeting rooms, work areas and catering areas are used. Our aim is to create a floor heatmap that visualizes crowd density from live, historical or simulated data as shown in Figure 5. In practice, each area may be covered by one or multiple APs. Based on geospatial information of APs, 1-to-N area-to-AP mapping can be defined to concatenate statistics associated with multiple APs to estimate the crowd density in each area. For instance, the number of visitors in a large meeting room equipped with two APs can be derived based on the number of unique devices (hashed MAC addresses) or unique users (hashed user IDs) connected to the corresponding APs. Such mappings are intent-based and not exclusive. Different mappings can be constructed for different analysis purposes and an AP can be associated to different mapping schemes. For instance, a COVID-19 purposed mapping may cluster APs by confined areas. For a health and wellbeing application that encourages people to use stairs rather than elevators, APs can be clustered based on their proximity to stairways or elevators to help understand the use of facilities. In the occasion of multiple areas sharing the same AP, previous locations of the connected devices can be used to estimate which area each user is at. Mappings are stored in a relational database as records of *Theme ID*, *WiFi AP ID*, *area code*, *area type*.

For crowd density analysis, it is important to differentiate stationary crowd (people who are staying in an area and purposely using the space) and crowd traffic (people who are moving across one area to access a different area). A person may pass by multiple areas before reaching her destination. This can leave a trace of "footprint" when her user device joins and leaves multiple APs on her path. In an education institution or business organisation with timetabled events, crowd traffic can generate significant "noises" in the crowd data. For instance, a group of 50 students leaving the classroom and exiting the building would cause a major fluctuation in crowd density measurement in all areas between the classroom and the exits of the building. For our use case, especially in the context of COVID-19, the analysis is centered around stationary crowd. Hence, we filtered out presence data that were observed for less than one minute by any AP. This was done by comparing every two consecutive samples and kept only the data records that appeared in both samples.

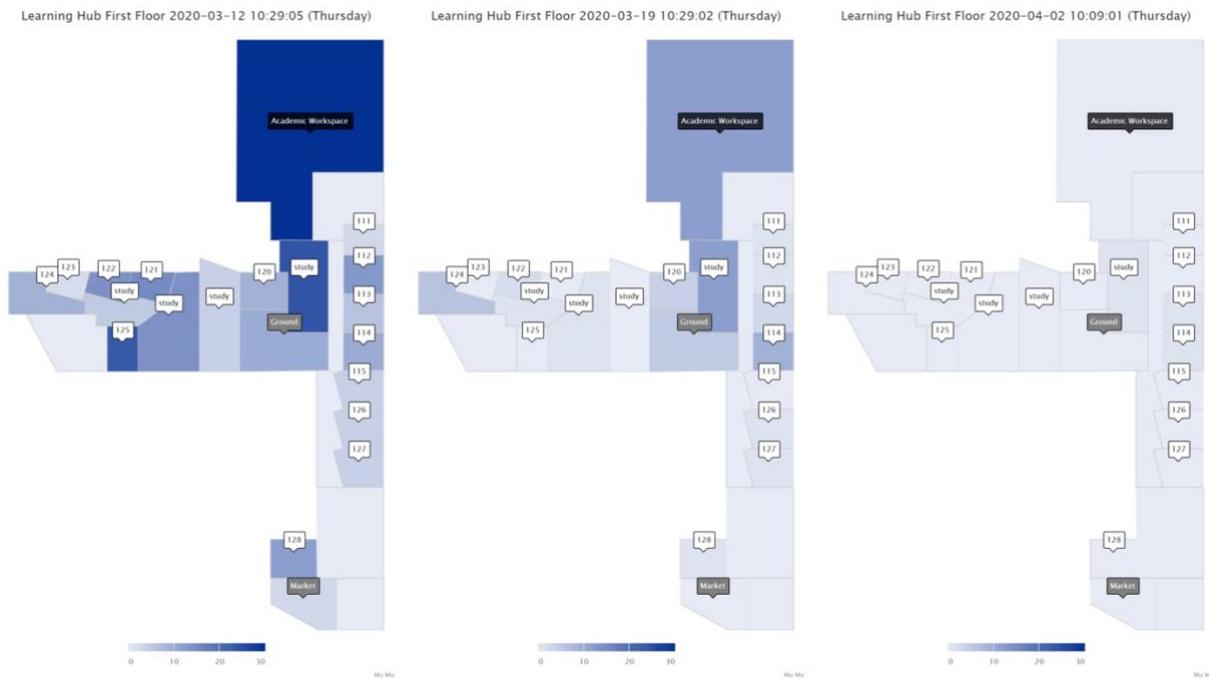

*Figure 5 A comparison of area crowd density of one floor of a building. from left to right: Pre-lockdown baseline, Near-lockdown reduced social contact, and working/studying from home during lockdown*

Figure 5 shows the crowd density heatmap of one building floor at three different points in time. All of our diagrams were developed based on the actual floors plan provided by the architect. This particular floor has eighteen lecture rooms and study areas (white label), one large academic workspace as an open-plan office (black label), and two catering areas (grey label). A deeper shade of blue color indicates a higher number of occupants in the area. The data is based on a 10-minute average of the number observed minute-by-minute. As people often have more than one device connected to the WiFi network (in most cases a smartphone and a laptop), we used unique user ID to measure the number of unique occupants. The heatmap is part of a web application that automatically updates every 10 minutes to show the live crowd density. It is also able to "replay" historical data at a configured playback speed.

The comparison between the three diagrams in Figure 5 provides a different view of the impact of COVID-19 lockdown. The heatmap on the left shows the data observed on Thursday March 12, 2020 at 10:29am which is part of a week not significantly affected by COVID-19. People were still using meeting rooms and study areas actively. The heatmap in the middle shows the data captured one week later at the same time when many people start to work from home voluntarily. Some people still chose to access the shared space, but the use of confined spaces especially small meeting rooms dropped significantly. Open areas became popular among visitors. The last heatmap shows a deserted building floor during the lockdown period. Only a small number of authorized staff had access to the building for maintenance and security check.

Crowd density heatmaps are intuitive to read. When augmented with overlay information, they can play a key role in helping visitors to gain confidence to return to work and to discover less populated areas to work. The heatmaps are to appear on large-format public displays that are strategically placed across campus for visitors. The information is also accessible on user devices as an interactive web application.

## Conclusions

The recovery from COVID-19 pandemic will take a long and cautious process. Informed risk management and policy making play a central role in mitigating the impact of COVID-19 and lessen the possibility of new outbreaks emerging. Capitalizing on a smart campus project and an experimentation environment, a crowd monitoring solution is developed to enable live monitoring, risk assessment, and planning of access to physical facilities in the context of COVID-19. The solution uses distributed wireless APs of an SDN-assisted network infrastructure as always-on IoT sensors to capture crowd data anonymously. Through data processing and visualization, crowd density and movements can be evaluated at different granularity levels. Predictive models are also developed to enable simulation and anomaly detection. The project now continuously evolves with additional COVID-19 related metrics and features such as contact tracing being developed. Data from different sensors such as proximity sensors are also considered to improve data accuracy in key areas and as a "method triangulation" tool.

## Acknowledgment

This work was supported by UK Research and Innovation (UKRI) under the EPSRC Grant Software Defined Cognitive Networking (EP/P033202/1). The author would like to thank Professor Nick Petford and Chris Foward for their valuable support within the smart campus project.